\documentclass{llncs}
\usepackage{graphicx}
\graphicspath{{./images/}}
\usepackage{csquotes}
\usepackage[table,xcdraw]{xcolor}
\usepackage{listings}
\usepackage{booktabs}
\usepackage{enumerate}
\usepackage[shortlabels]{enumitem}
\usepackage[T1]{fontenc}
\usepackage{amsmath,algorithm,algpseudocode}
\usepackage{multirow}
\usepackage{makecell}

\usepackage{todonotes}

\begin{document}
\title{ConTrOn: Continuously Trained Ontology based on Technical Data Sheets and Wikidata}
\author{Kobkaew~Opasjumruskit\inst{}\orcidID{0000-0002-9206-6896}\and 
Diana~Peters\inst{} \orcidID{0000-0002-5855-2989}\and 
Sirko~Schindler\inst{}\orcidID{0000-0002-0964-4457}}
\institute{DLR Institute of Data Science\\
M{\"a}lzerstra{\ss}e 3, 07745 Jena, Germany\\
\email{\{firstname.lastname\}@dlr.de}}

\maketitle
\pagestyle{headings}
\begin{abstract}
In engineering projects involving various parts from global suppliers, one common task is to determine which parts are best suited for the project requirements.
Information about specific parts' characteristics is published in so called data sheets.
However, these data sheets are oftentimes only published in textual form, e.g., as a PDF. 
Hence, they have to be transformed into a machine-interpretable format.
This transformation process still requires a lot of manual intervention and is prone to errors.
Automated approaches make use of ontologies to capture the given domain and thus improve automated information extraction from the data sheets.
However, ontologies rely solely on experiences and perspectives of their creators at the time of creation and cannot accumulate knowledge over time on their own. 
This paper presents ConTrOn -- Continuously Trained Ontology -- a system that automatically augments ontologies. 
ConTrOn tackles terminology problems by combining the knowledge extracted from data sheets with an ontology created by domain experts and external knowledge bases such as WordNet and Wikidata. 
To demonstrate how the enriched ontology can improve the information extraction process, we selected data sheets from spacecraft development as a use case.
The evaluation results show that the amount of information extracted from data sheets based on ontologies is significantly increased after the ontology enrichment.

\keywords{Ontology enrichment $\cdot$ Ontology-based information extraction $\cdot$ Knowledge representation}
\end{abstract}

\section{Introduction}

In the development and assembly process of most engineering projects technical data sheets are a major source of information about the components an engineer might want to use.
There is, especially in the space domain, no single source for data sheets, but engineers have to get them directly from the manufacturers or sometimes from shops and distributors, as mentioned by Peters et al. \cite{DP2019}. 
Because of this, the selection of data sheets is mainly based on personal experience and preferences and it would broaden the area of possible components if this selection could be based on technical requirements as well.
After the selection of a data sheet comes the extraction of the contained information in other planning or design tools.
Since PDF is not machine-understandable, this has to be done by hand, which is not only time and energy consuming, but also error-prone.

The aim of ConTrOn (Continuously Trained Ontology) is to extract the information from PDF data sheets and make it available in a machine-understandable format.
This allows also to search for information more easily and can therefore lead to the discovery of components an engineer did not know about before.
Since the problem emerged for us in the space domain, our examples are located there, but we see several possibilities for generalization.
First, our approach can be applied to PDF data sheets of other technical disciplines as well since the formats are quite similar.
Second, our approach can be applied to other structured documents as well, e.g. medical statements and financial documents.
A prerequisite for both types of adaption is a domain specific ontology.
Our approach helps then to improve this ontology, which must be initialized by domain experts.

For ConTrOn, we first collected implicit knowledge from domain experts regarding relevant properties of spacecraft parts, such as volume, mass, and radiation tolerance and represented it as ontologies.
Then we used an information extraction process on domain specific data sheets which led to new terms and relations that were then added to the ontologies.
These enriched ontologies then helped to improve the information extraction process and allowed the discovery of more information out of the unstructured text.

Following this workflow, we state our research questions as follows:
\begin{description}
  \item[R1:] How can essential data, such as physical attributes of each component, be automatically extracted  from unstructured data sheets?
  \item[R2:] How can the essential data be presented in a form of ontology?
  \item[R3:] How can one get semantic knowledge to enrich the ontology?
\end{description}

ConTrOn is a system to answer these questions, using existing Semantic Web techniques like PDF Information Extraction, Ontology Learning, and Ontology-Based Information Extraction.
Based on classes defined by domain experts in the initial ontologies, we extract textual information from data sheets.
The extracted texts are used to retrieve explicit knowledge from external data sources, such as Wikidata \cite{Vrandecic2014}
or WordNet \cite{Fellbaum2000}.
We use those general sources since we are not aware of space domain specific ones and of the purposes of ConTrOn is the generation of a a domain specific knowledge source (i.e. the enhanced ontology).
However, keywords used in retrieving external information can be ambiguous.
Natural Language Processing (NLP) can be used to help analyze the context of data sheets and choose the best-matched definition of the keyword.
We then use the concepts retrieved from external knowledge bases to extend our ontologies.
We iterate the process as soon as new data sheets are available to automatically enrich the ontologies over time. 
Furthermore, compared to the keyword-based information extraction, ontologies provide more relevant concepts and thus increase the amount of discovered information.

In the next section, we provide background knowledge for our proposed solution as well as a review of the existing literature. 
In Section \ref{sec:solution}, we elaborate on the solution workflow and provide details of each workflow module. 
Each module was evaluated based on the evaluation setups described in Section \ref{sec:evaluationSettings} and the results are shown and discussed in Section \ref{sec:evaluationResults}.
Finally, the conclusion of this paper and the ideas for future work are proposed in Section \ref{sec:conclusion}.

\section{Related Work}\label{sec:relatedWork}
In ConTrOn, we are mainly dealing with the following tasks: PDF data extraction, ontology learning (OL) and ontology-based information extraction (OBIE). 
We base our review of existing approaches on this categorization and provide individual reviews in the followings subsections.

\subsection{PDF information extraction and processing}
The PDF information extraction, including text pre-processing, is the most mature among the three tasks in terms of availability of tools.
The text extraction from PDF files has been tackled by both commercial, such as pdf2Data\footnote{\url{https://itextpdf.com/en/products/itext-7/pdf2data}} and PDF tables\footnote{\url{https://pdftables.com/}}, and open source solutions.
Tabula \footnote{https://tabula.technology/} is a free tool to extract tabular data from PDF files.
Robotips' uConfig tool \footnote{\url{https://github.com/Robotips/uConfig}} is a specific solution to extract schematic parts and footprints.
Issues such as a multiple columns layout mixed with text, graphs, and tables can be solved using layout-aware text extraction techniques as proposed by Chao and Fan \cite{Chao2004} or Ramakrishnan et al. \cite{Ramakrishnan2012}.

In technical data sheets, text, table layouts, and schematic drawings are equally important.
However, most of the existing tools focus on either text or tables and there is no unified solution which tackles all of these.

\subsection{Ontology Learning}
Semi-automatic approaches for ontology learning have been proposed since the end of the 90's \cite{Petasis2011}.
However, this research topic has been repeatedly updated due to the availability of new technologies and resources. 
Most of them use NLP techniques to detect entities, extract information and relations from plain text documents or web pages, such as TEXT-TO-ONTO \cite{Maedche2000} and KAON \cite{Bozsak2002}.
Some approaches also semantically and syntactically analyze external knowledge bases, such as WordNet or web pages, to enrich their ontologies \cite{Agirre2000,Moldovan2000}.

Asim et al. \cite{Asim2018} categorized ontology learning into three techniques: linguistic, statistical, and logical.
Linguistic techniques exploit characteristics of the used language, statistical techniques are based on a statistical analysis of the text corpora, and logical techniques resort to machine learning (ML) approaches.
There is no clear-cut conclusion which technique is better, since their performance relies on the domain and amount of documents they use.

\subsection{Ontology-Based Information Extraction}\label{sec:relatedWork:obie}
Baclawski et al. \cite{Baclawski2017} summarized the current research tracks that combine ML, information extraction and ontology techniques to solve complex problems, such as OBIE.
OBIE, as described by Wimalasuriya and Dou \cite{Wimalasuriya2010}, is a system that processes unstructured or semi-structured text to extract certain types of information guided by ontologies and present the output as instances of those ontologies.
The extracted information from an OBIE system is used not only to populate and enrich ontologies, but also to improve NLP workflows.

Maynard et al. \cite{Maynard2008} described NLP techniques for ontology population using OBIE.
XONTO \cite{Oro2008} proposed an OBIE system for semantic extraction of data from PDF documents with the guide of ontologies.
In contrast to XONTO, Dal and Maria \cite{Dal2012} suggested an ontology creation method using ML and external knowledge. 
They extract concepts from documents using latent semantic analysis and clustering techniques. 
Meanwhile, properties, axioms, and restrictions are retrieved from WordNet.

Barkschat \cite{Barkschat2014} proposed an OBIE workflow that exploits technical data sheets to populate ontologies using a classifier model and regular expressions.
Likewise, Smart-dog \cite{Murdaca2018} extracts data from data sheets of spacecraft parts to populate an ontology.
It features an ontology enrichment step, but relies on domain experts. 
Meanwhile, Rizvi et al. \cite{Rizvi2018} included irrelevant terms and probably-relevant terms in their ontology so that they can calculate the confidence score of the extracted information.

\section{ConTrOn Workflow}\label{sec:solution}
\begin{figure}[!htb]
  \includegraphics[width=\columnwidth]{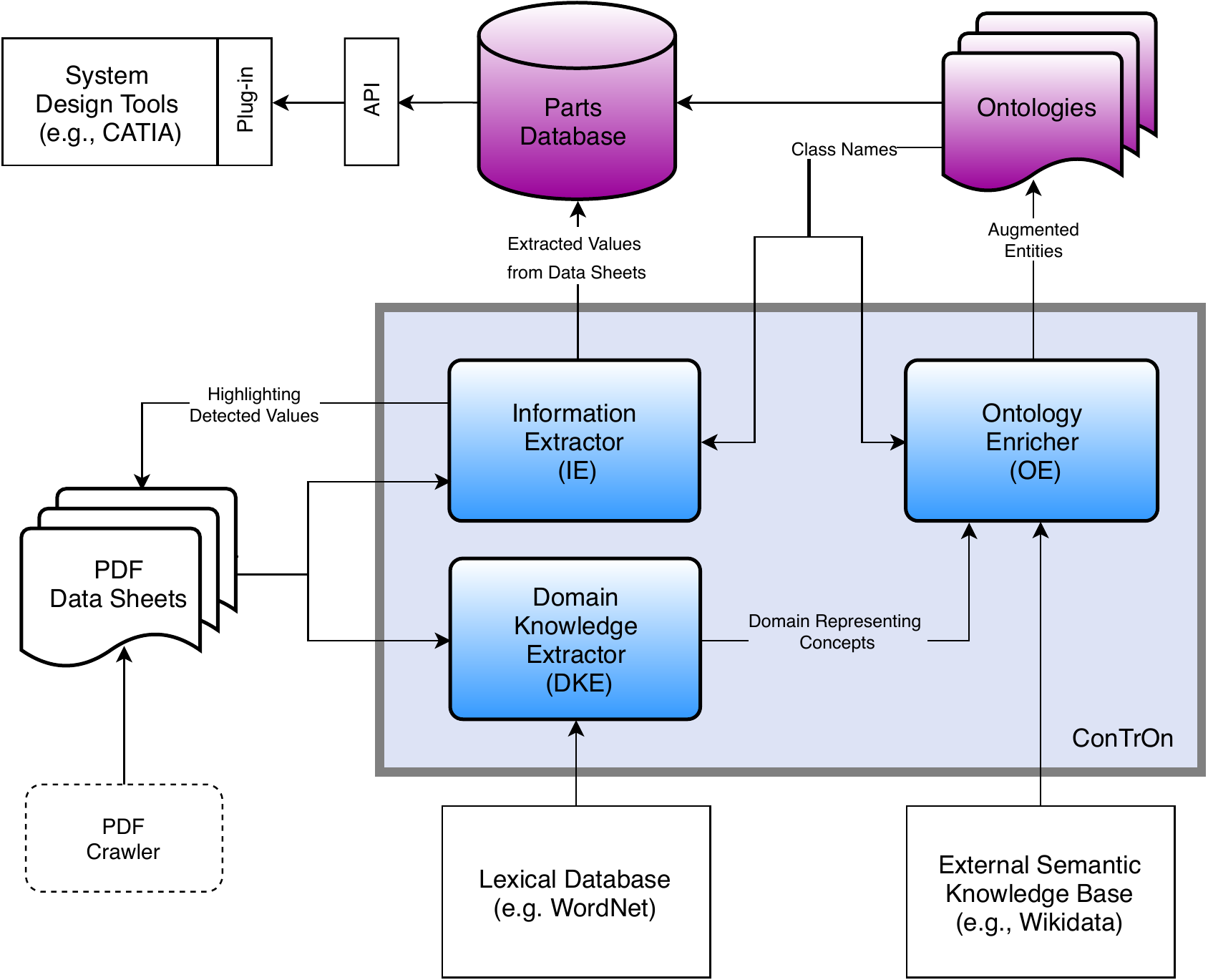}
  \centering
  \caption{ConTrOn Architecture.}
  \label{fig:block-diagram}
\end{figure}

ConTrOn offers a solution that enriches ontologies using semantic knowledge bases and extracts information from data sheets guided by the enriched ontology. 
This section starts with an overview of the system, as shown in Fig. \ref{fig:block-diagram}.
Each of its main modules, Domain Knowledge Extractor (DKE), Ontology Enricher (OE), and Information Extractor (IE), will be explained in detail afterwards.

\subsection{System Overview}
ConTrOn uses data sheets sampled from various manufacturers' websites. 
The workflow starts by reading the text from the collected data sheets. 
DKE extracts a set of \emph{domain representing concepts}. 
It is very likely that multiple definitions are found for each concept. 
To disambiguate the meaning of these concepts (word-sense disambiguation), DKE analyzes the definitions provided by WordNet.
This process can be scheduled to run each time a new data sheet is available. 

Afterwards, OE retrieves class names from the initial ontologies prepared by domain experts. 
OE retrieves semantic definitions of ontologies' classes from external resources, such as Wikidata.
For word-sense disambiguation, OE compares the definitions with the domain representatives computed by DKE.
The best matched definition of each class is then extracted for semantic description and is used to augment the class.
Classes in the enriched ontologies contain \enquote{alternative labels}, \enquote{synonyms}, and \enquote{categories}.

IE searches in data sheets for text surrounding given \emph{keywords} to retrieve corresponding values. 
The \emph{keywords} are not limited only to a single label per concept, but can also be the additional labels acquired from OE.
The discovered \emph{keywords} and their values are highlighted within the data sheets for later reference.
Finally, the extracted information will be exported to engineering applications for further usage.

\subsection{Domain Knowlege Extractor (DKE)}\label{sub:ie}

DKE analyzes texts from data sheets and extracts words that represent concepts of the text corpus, e.g., \enquote{telemetry} and \enquote{payload} (see Algorithm \ref{alg:dke}). 
Initially, text from all data sheets is tokenized into a group of words (bag-of-words), regardless word sequences. 
In the tokenization, DKE considers not only unigrams (terms with one word), but also n-grams (terms that contains multiple words, 
such as \enquote{magnetic field} or \enquote{propulsion system}) to retain their concepts.

\begin{algorithm}[!htb]
\caption{Domain Knowledge Extractor (DKE) Algorithm.}
\label{alg:dke}
\begin{algorithmic}[1]\small
\Function{DomainKnowledgeExtractor}{$text$}
    \State $terms$ = tokenized $text$
    \State $topics$ = get $terms$ with high Tf-idf
    \State get $synsets$ of each $topic$ from WordNet
    \State $similarity\_matrix$ = semantic similarities between each pair of $synsets$ 
    \For{$topic$ in $topics$}
       \State $concepts$ += $synset$ of the topic with the highest similarity value
    \EndFor
    \State \Return $concepts$
\EndFunction
\end{algorithmic}
\end{algorithm}

Next, DKE computes the TF-IDF score for each term.
Terms with a high score are considered domain topics.
As topics may have multiple meanings, we employ WordNet\footnote{\url{https://wordnet.princeton.edu/}} for disambiguation.
WordNet is a lexical database for the English language and is used to define words with regard to the context of usage. 
Each word is defined by one or more synsets consisting of a \textit{unique id}, \textit{definition}, \textit{example of usage}, and relevant terms such as \textit{synonyms} and \textit{hypernyms}. 

To find the best matching definition, DKE first retrieves all corresponding synsets for a given topic.
Then DKE calculates a graph using synsets as vertices and the respective semantic similarity\footnote{
Similarity calculations are performed using Wu-Palmer Similarity \cite{Wu1994} provided by Python's Natural Language Toolkit (NLTK)\footnote{\url{http://www.nltk.org/}}.} 
between them as a weight to the edges.
Note that there are no similarities calculated within the synsets of one topic, as this would not contribute to the goal of selecting just one of them.
For each synset, DKE sums up the weights of all adjacent edges.
Then, for each topic, DKE selects the synset with the highest accumulated score.
This results in a collection of synsets that are most relevant among each other; thus, most likely representing the meaning within the data sheet.

\begin{figure}
  \includegraphics[width=0.95\columnwidth]{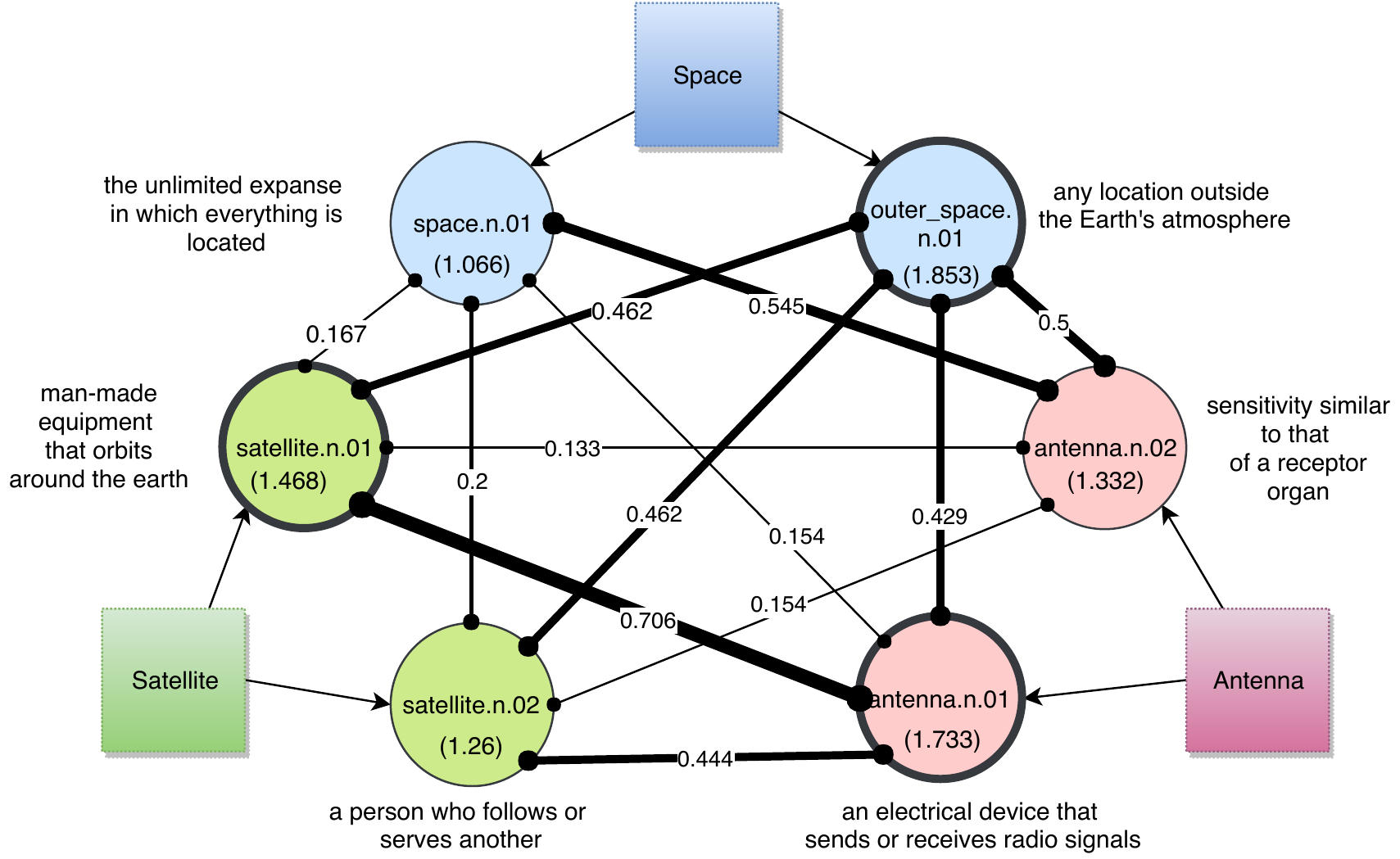}
  \centering
  \caption{Similarity between synsets (circle shapes) is used to disambiguate the meaning of topics (square shapes). Edges represent the similarity between them with a weight $0$ referring to irrelevant and $1$ to semantically identical. The sum over all adjacent edges is given for each synset in parenthesis.}
  \label{fig:synsets_similarities}
\end{figure}

Consider Fig. \ref{fig:synsets_similarities} as an example.
In this case, three topics have been extracted from the data sheets (in square shapes): \enquote{space}, \enquote{satellite}, and \enquote{antenna}.
Each of them is associated with synsets (in circle shapes).
The value on each edge represents the semantic similarity between two adjacent vertices. 
For instance, the similarity between \enquote{outer\_space.n.01} and \enquote{antenna.n.01} is $0.429$.
Next, the sums of adjacent edge weights is calculated for each vertex, shown in parenthesis within each circle.
To select the preferred synset, for each topic the vertex/synset with the highest accumulated weight is chosen.
Regarding the topic \enquote{space}, the decision is between \enquote{space.n.01} and \enquote{outer\_space.n.01}. 
The latter vertex/synset is favored with a score of $1.853$ over the former's score of $1.066$.

Finally, DKE returns a set of concepts with their definitions from WordNet synsets.
These concepts form the \emph{essential data} of the data sheets, which correspond to research question R2.

\subsection{Ontology Enricher (OE)}

To address research question R3, OE retrieves data from external knowledge bases to enrich an existing ontology (see Algorithm \ref{alg:oe}). 
Classes from the ontology may lack a description, related concepts, and alternative names, which can be retrieved from a semantic knowledge base like Wikidata.

\begin{algorithm}[!htb]
\caption{Ontology Enricher (OE) Algorithm.}
\label{alg:oe}
\begin{algorithmic}[1]\small
\Function{OntologyEnricher}{$ontology$, $concepts$} 
    \State $classes$ = read classes from $ontology$
    \For{$class$ in $classes$}
       \State get $entities$ corresponding to $class$ name from Wikidata	        
       \State $similarity\_matrix$ = cosine similarities between $entities$  and $concepts$
 	 \State $matched\_entity$ = entity with the highest similarity value
 	 \State update $class$ with $matched\_entity$
    \EndFor 
    \State $enriched\_ontology$ = $ontology$ with updated $classes$
    \State \Return $enriched\_ontology$
\EndFunction
\end{algorithmic}
\end{algorithm}

Very often, a class name matches multiple entities or has no match at all.
For the first issue, OE disambiguates the class name by comparing each matched entity description with the domain representing concepts obtained by DKE.
Each entity description is considered a document, while the domain representing concepts are also considered another document.
Assuming that we obtained $n$ terms from all documents, OE computes a Vector Space Model (VSM) where each document is represented as a vector in an n-dimensional space.
The cosine similarity between entity-vectors and the vector representing domain concepts is used to determine which entity should be used to augment the ontology.

It can be argued that the augmentation should be done automatically.
To prevent semantic drift caused by misconception, we defined a confidence threshold to ensure that the selected entity has a certain level of similarity.
If the similarity between the best matching entity and the domain knowledge is above the threshold value, OE assumes that the entity represents the ontology class.
On the one hand, if there is one entity that achieve the criterion, OE automatically augments the ontology with this entity.
On the other hand, if no entity meets the requirement, or multiple entities meet the requirements, OE will prompt the domain experts to review the related entities.

If no matching entity is found, OE follows the intuitive approach by retrieving synonyms and relevant terms of the original keywords from WordNet.

\subsection{Information Extractor (IE)}

In the previous subsections, we explained how to extend ontologies with external knowledge.
Now, these enriched ontologies can be used to discover more information from data sheets 
(see Algorithm \ref{alg:ie}), and thus to address research question R1.
In addition to the original class names, IE also uses alternative keywords, such as \enquote{label}, \enquote{alternative label}, \enquote{category} from the enriched ontology.

\begin{algorithm}[!htb]
\caption{Information Extractor (IE) Algorithm.}
\label{alg:ie}
\begin{algorithmic}[1]\small
\Function{InformationExtractor}{$enriched\_ontology, text$} 
    \State get $classes$ from $enriched\_ontology$
    \For{$class$ in $classes$}
        \State $keywords$ = $class$ name, alternative labels, category
    	  \State $key\_value$ = corresponding values of $keywords$ in $text$
        \State $extracted\_information$ += $key\_value$ 
    \EndFor 
    \State \Return $extracted\_information$
\EndFunction
\end{algorithmic}
\end{algorithm}

Currently, the IE does the text-based search for the keywords, then looks around the keywords for corresponding values since most of the keywords are followed by their corresponding value, such as \enquote{temperature 40$^\circ$c} or \enquote{frequency 4 Hz}.
If the keywords are not followed by numeric values, the IE will perform a part-of-speech tagging over the surrounding text.
Then, the IE will search the tagged text for patterns defined as regular expressions, such as a sentence pattern (full stop + space + a capital letter or full stop + end of line) or a list pattern (two consecutive lines that start with capital letters).

If the extracted text contains numeric values, we assume that these numbers are the corresponding values and keep them for later use.
Otherwise, we rely on the engineers or domain experts to review the extracted text fragment and select the respective value.
Discovered keywords and their surrounding text will be highlighted in the data sheets and annotated with a reason for the highlighting, e.g., \enquote{The highlighted text (Life span: 5 Years) is corresponding to the Lifetime property}.

\section{Evaluation Setup}\label{sec:evaluationSettings}
In this section we describe the prerequisites used for our evaluation.
Kindly recall, that our proposed workflow requires three kinds of inputs:
a set of data sheets, an initial ontology, and a set of external knowledge bases.
We employed the publicly available endpoints of Wordnet and Wikidata for the latter.
In the following, we will describe the ontologies as well as the data sheets used.
We chose spacecraft manufacturing as a use case for this evaluation.
%The source code of ConTrOn used in this evaluation is published on \cite{contronCode2019}.

\textbf{Ontology:}
The first, general ontology was prepared based on satellite component data derived from data models from a Model-Based Systems Engineering tool, Virtual Satellite \cite{Fischer2017}, and feedback from domain experts.
This ontology covers concepts common to most satellite parts, thus forming a core ontology.
The second, specialized ontology was created to describe a specific satellite category, the star tracker.
Both ontologies are publicly available from \cite{contron2019}.

\textbf{Corpus:}\label{sub:textCorpus}
From an online search, we obtained 170 data sheets describing more than three hundred satellite parts\footnote{We consider this a random sample from a population of unknown size, since there is no single source of satellite parts data sheets \cite{DP2019}.}.
The numbers of data sheets and products differ because one data sheet can contain multiple products.
All data sheets are used to evaluate the core ontology, while the star tracker data sheets are used to evaluate the enriched star tracker ontology.
In the following we give a summary over the retrieved data sheets.

\begin{multicols}{2}
\begin{itemize}\scriptsize
\item Battery (6 files, 14 products)
\item Camera (8 files, 18 products)
\item Cubesat (1 file, 1 product)
\item Earth sensor (5 files, 5 products)
\item GNSS (14 files, 17 products)
\item GNSS receiver (16 files, 16 products)
\item Magnetometer (13 files, 14 products)
\item Magnetic torquer (7 files, 12 products)
\item Pocketqube (3 files, 3 products)
\item Reaction momentum wheel	(20 files, 33 products)
\item S-band (14 files, 17 products)
\item \textbf{Star tracker (16 files, 19 products)}
\item Sun sensor (10 files, 16 products)
\item Thruster (26 files, 87 products)
\item X-band (11 files, 11 products)
\end{itemize}
\end{multicols}

The acquired data sheets were randomly split into three partitions (Partition I to III).
The test-corpora are built by iteratively adding those partitions to the analyzed set of data sheets:
Corpus A consists of just Partition I and its $38$ files.
Corpus B also includes Partition II resulting in $38 + 61 = 99$ files.
Finally, Corpus C includes all partitions and thus $38 + 61 + 71 = 170$ files.
This approach resembles the envisioned workflow where we expect new batches of data sheets occasionally being released and analyzed.

\newenvironment{conditions}
  {\par\vspace{\abovedisplayskip}\noindent\begin{tabular}{>{$}l<{$} @{${}={}$} l}}
  {\end{tabular}\par\vspace{\belowdisplayskip}}
  
\subsection{Evaluation Metrics}
According to Maynard et al. \cite{Maynard2006}, traditional metrics for evaluating information extraction systems are precision, recall and F-measure.
Precision represents the ratio between the number of correctly identified items (True Positive) and the number of items identified (True Positive+False Positive). 
Recall represents the ratio between the number of correctly identified items and the total number of correct items (True Positive+False Negative). 
F-measure is a harmonic average of precision and recall. 
\begin{equation}\label{eqn:fmeasure}
F-measure = \frac{(\beta^2 +1) \cdot precision\cdot recall}{precision+ (\beta^2 \cdot recall)}
\end{equation}
where:

\begin{conditions}
precision & ${True~Positive} ~/~ (True~Positive+False~Positive)$\\
recall    & ${True~Positive} ~/~ (True~Positive+False~Negative)$\\
\beta     & The priority of precision over recall
\end{conditions}

In our evaluation, we chose $\beta$ as 1 because we regard precision and recall as equally important in this context.

\section{Evaluation Results}\label{sec:evaluationResults}

The evaluation examines the effects of ConTrOn on the information extraction process.
We split the evaluation into three part according to the modules.
Since there is no publicly available tool that performs the OBIE task on PDF files, we selected a plain text search as a baseline.
The keywords used are based on the class-names of the initial ontologies.

The full set of acquired data sheets (170) is only used for the evaluation of the DKE and the OE.
Here, only a limited number of results is subject to manual review.
However, the result set of the IE is much larger and thus prevents the use of all data sheets here.
We selected a subset of the data sheets -- in particular, all 16 data sheets referring to star trackers -- for this final part of the evaluation.

\subsection{Domain Knowledge Extractor}

For the full set of 170 data sheets DKE had to consider about 208 thousand words.
Removing specific names and stop words lowered this number to about 85 thousand words.
Out of those, around 50 thousand words could be mapped to 5203 unique topics in WordNet.
Applying the TF-IDF threshold leaves us 680 domain topics for Corpus C.
The same process results in 320 domain topics for Corpus A and 600 for Corpus B.

Finally, the VSM is applied to disambiguate the extracted topics and thus determine the essential data for the given data sheets.
Table \ref{tab:ldatopics} lists the five most significant topics for each corpus alongside the respective matched synset id.

\begin{table}[h]
\renewcommand{\arraystretch}{1.3}
\setlength\tabcolsep{6pt}
\resizebox{\textwidth}{!}{%
\begin{tabular}{ l | l | l | l }
\hline
\rowcolor{gray!50}
\multicolumn{1}{c|}{\cellcolor{gray!50}Rank} 
& \multicolumn{1}{c|}{\cellcolor{gray!50} Corpus A} 
& \multicolumn{1}{c|}{\cellcolor{gray!50} Corpus B} 
& \multicolumn{1}{c}{\cellcolor{gray!50} Corpus C} \\
\hline
	1& false(false.a.01) & power(ability.n.02) & data(datum.n.01)\\

	\rowcolor{gray!20}
	2& true(true.a.01) & data(datum.n.01) & power(ability.n.02)\\

	3 & up\_to(up\_to.s.01) & output(output\_signal.n.01) & up\_to(up\_to.s.01)\\
	
	\rowcolor{gray!20}
	4 & https(hypertext\_transfer\_protocol.n.01) & time(time.n.05) & output(output\_signal.n.01)\\

	5 & power(ability.n.02)  & performance(performance.n.03) & system(system.n.02)\\
	
\hline
\end{tabular}
}\\

\caption{The five most significant topics and the corresponding concepts (denoted by WordNet' synset ids) based on the Corpus A, B, and C.} \label{tab:ldatopics}
\end{table}

While we see rather generic terms ranked quite highly for the small Corpus A, their influence diminishes for a larger Corpus B and C.
When comparing the results for Corpus B and C, we recognize the early effects of saturation with respect to extracted topics:
The vocabulary used within this data sheets of a specific domain is fairly limited, so we expect the distribution of topics to stabilize given a sufficient number of data sheets.
From this observation, we expect only a minor improvements of the DKE for input sizes beyond the examined one.

\subsection{Ontology Enricher}\label{sub:oe-eval}
Initially, both ontologies contain only information added by domain experts.
During the enrichment process, their classes are augmented with explicit information retrieved from Wikidata. 

\subsubsection{Core Ontology:}
First, we evaluated how the amount of data sheets affect the OE process. 
Using the core ontology and the topics obtained from the DKE based on the Corpus A, B, and C, the OE found matching entities for 26 classes as summarized in Table \ref{tab:ont-enriched-core-no-iteration}.
%The confidence threshold was empirically configured to 0.6 - indicating that the best result must have a similarity value compared to the domain representatives at least 60\% above the mean similarities of all results.

%\todo{KO: Split the table and figure.}
\begin{table}[h]

  \renewcommand\cellgape{\Gape[4pt]}
  \renewcommand{\arraystretch}{1.3}
  \setlength\tabcolsep{6pt}

  \resizebox{\textwidth}{!}{%
  \begin{tabular}{ c | c c c c | c c c }
    \hline
    \multicolumn{1}{c|}{ } 
    & \multicolumn{2}{c|}{ \makecell{Classes with\\ one entity \\assigned} } 
    & \multicolumn{2}{c|}{ \makecell{Classes with\\ multiple/no entities\\assigned} } 
    & & & \\
    
    \cline{2-5}
    
      \multirow{-2}{*}{Corpus}
    & \makecell{ I   \\ correct }
    & \makecell{ II  \\ incorrect }
    & \makecell{ III \\ correct }
    & \makecell{ IV  \\ incorrect }
    & \multirow{-2}{*}{Recall}
    & \multirow{-2}{*}{Precision}
    & \multirow{-2}{*}{F-measure} \\
    \hline
    	\rowcolor{gray!20}
    	A & 8  & 5 & 8 & 5
    	& 0.38 & 0.31 & 0.34 \\
    	%& $\frac{8}{21}=0.38$ & $\frac{8}{26}=0.31$ & 0.34 \\
    
    	B& 5 & 5 & 14 & 2 
    	& 0.21 & 0.19 & 0.20\\
    	%& $\frac{17}{25}=0.68$ & $\frac{17}{26}=0.65$ & 0.67\\
    
    	\rowcolor{gray!20}
    	C & 6 & 4 & 13 & 3 
    	& 0.26 & 0.23 & 0.24\\
    	%& $\frac{23}{24}=0.96$ & $\frac{23}{26}=0.88$ & 0.92\\
    	
    %\hline
    %	&A& 6  & 4 & 11 & 8
    %	& 0.29 	& 0.21 & 0.24 \\
    %	%$\frac{6}{21}=0.29$ 	& $\frac{6}{29}=0.21$ & 0.24 \\
    %
    %	\rowcolor{gray!20}
    %	\cellcolor{white}& B& 5 & 4 & 14 & 6 
    %	& 0.22 & 0.17 & 0.19\\
    %	%& $\frac{17}{22}=0.77$ & $\frac{17}{29}=0.59$ & 0.67\\
    %
    %	\multicolumn{1}{c|}{ \multirow{-3}{*}{\begin{tabular}[c]{@{}c@{}}
    %	Star \\Tracker\end{tabular} }} 
    %	& C& 6 & 3 & 12 & 8 
    %	& 0.29 & 0.21 & 0.24\\
    %	%& $\frac{21}{21}=1$ & $\frac{21}{29}=0.72$ & 0.84\\
    \hline
  \end{tabular}
}\\

\includegraphics[width=0.99\columnwidth]{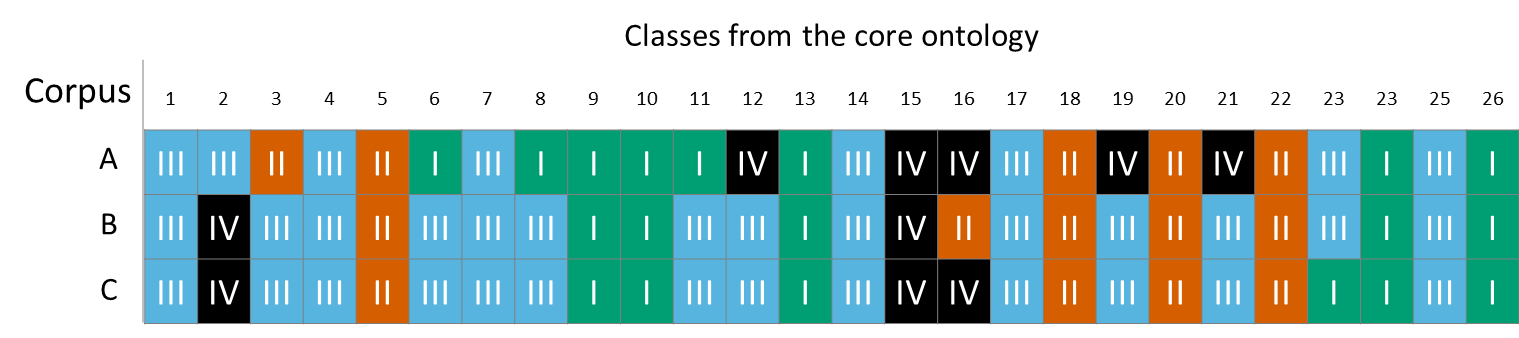}

\caption{Classes in the core ontology were matched to relevant entities by the OE with the various size of data sheets.} \label{tab:ont-enriched-core-no-iteration}
\end{table}

Looking at the number of correct results (group I) alone, the OE cannot perform better by including more data sheets.
Take the class number 2 in Table \ref{tab:ont-enriched-core-no-iteration} for example, the increment in topics number introduced noises and led to a misconception.
However, when looking at other classes, e.g. number 3, 12, and 19, the bigger the Corpus is, the more relevant entities are discovered. 
The benefit of increasing the amount of data sheets lies in the entities that are waiting to be selected by human.

Therefore, we evaluated the OE further by iteratively exectuting the OE and having a domain expert to edit the ontology after each iteration.
Incorrect entities (group II) that were assigned to classes were disjointed so that the OE can avoid mismatching in the future.
Classes in the group III were reviewed and the matching entities were selected.
After the review process, the edited ontology was used again as an input to the OE in the next iteration.
The results of this iterative evaluation are shown in Table \ref{tab:ont-enriched}.

\begin{table}[h]\small
\renewcommand{\arraystretch}{1.3}
\setlength\tabcolsep{6pt}
\resizebox{\textwidth}{!}{%
\begin{tabular}{ c | c | c c c c | c c c }
\hline
\multicolumn{1}{c|}{ } 
&\multicolumn{1}{c|}{ } 
& \multicolumn{2}{c|}{ \begin{tabular}[c]{@{}c@{}}
	Classes with\\ one entity \\assigned\end{tabular} } 
& \multicolumn{2}{c|}{ \begin{tabular}[c]{@{}c@{}}
	Classes with\\ multiple/no entities\\assigned\end{tabular} } 
& & &   \\ \cline{3-6}

\multicolumn{1}{c|}{ \multirow{-2}{*}{Ontology}} 
&\multicolumn{1}{c|}{ \multirow{-2}{*}{\begin{tabular}[c]{@{}c@{}}
      Iteration\\ /Corpus \end{tabular}}} 
& \multicolumn{1}{c|}{ \begin{tabular}[c]{@{}c@{}}I\\ correct \end{tabular} }  
& \multicolumn{1}{c|}{ \begin{tabular}[c]{@{}c@{}}II\\ incorrect \end{tabular}}  
& \multicolumn{1}{c|}{ \begin{tabular}[c]{@{}c@{}}
	III\\ correct \end{tabular} } 
& \multicolumn{1}{c|}{ \begin{tabular}[c]{@{}c@{}}
      IV\\ incorrect \end{tabular}} 
& \multicolumn{1}{c}{ \multirow{-2}{*}{Recall}}  
& \multicolumn{1}{c}{ \multirow{-2}{*}{Precision}}  
& \multicolumn{1}{c}{ \multirow{-2}{*}{F-measure}} \\
\hline
	\rowcolor{gray!20}
	&1 / A & 8  & 5 & 8 & 5
	& 0.38 & 0.31 & 0.34 \\
	%& $\frac{8}{21}=0.38$ & $\frac{8}{26}=0.31$ & 0.34 \\

	\cellcolor{gray!20}&2 / B & 17 & 2 & 7 & 1 
	& 0.68 & 0.65 & 0.67\\
	%& $\frac{17}{25}=0.68$ & $\frac{17}{26}=0.65$ & 0.67\\

	\rowcolor{gray!20}
	\multicolumn{1}{c|}{ \multirow{-3}{*}{Core}} 
	&3 / C & 23 & 0 & 1 & 2 
	& 0.96 & 0.88 & 0.92\\
	%& $\frac{23}{24}=0.96$ & $\frac{23}{26}=0.88$ & 0.92\\
	
\hline

	&1 / A & 6  & 4 & 11 & 8
	& 0.29 	& 0.21 & 0.24 \\
	%$\frac{6}{21}=0.29$ 	& $\frac{6}{29}=0.21$ & 0.24 \\

	\rowcolor{gray!20}
	\cellcolor{white}&2 / B & 17 & 2 & 3 & 7 
	& 0.77 & 0.59 & 0.67\\
	%& $\frac{17}{22}=0.77$ & $\frac{17}{29}=0.59$ & 0.67\\

	\multicolumn{1}{c|}{ \multirow{-3}{*}{\begin{tabular}[c]{@{}c@{}}
	Star \\Tracker\end{tabular} }} 
	&3 / C & 21 & 0 & 0 & 8 
	& 1 & 0.72 & 0.84\\
	%& $\frac{21}{21}=1$ & $\frac{21}{29}=0.72$ & 0.84\\
	
\hline
\end{tabular}
}\\

\caption{Number of classes in the core and star tracker ontologies enriched by the OE. The ontologies were incrementally enhanced after each iteration. } \label{tab:ont-enriched}
\end{table}

In the first iteration, the OE chose the best entities for thirteen classes with a degree of confidence.
However, the automatically selected entities have low precision, that is eight classes are defined correctly (Table \ref{tab:ont-enriched}, core ontology, group I) and incorrect entities are selected for five classes (group II).
The classes in the group II were assigned with wrong entities because they have ambiguous names.  
%Five classes, for which wrong entities were selected, are listed in Table \ref{tab:ont-enriched-core-false}.
For instance, the class \enquote{Hardware Interface} was matched with a Wikidata entity which has the same name but refers to a study on a design of physical property instead of the physical property itself.

The other thirteen classes (group III + IV) were provided with a list of possible entities because none of the entities are distinctive enough to be chosen as the best result.
For example, \emph{Interface} can be inferred to a physical interface or a software interface.
These general concepts appear in many different contexts and should therefore be reviewed by domain experts before being inserted into the ontology.

However, no relevant answer was found in the returned list for five classes (group IV), such as \emph{Radiation Tolerance} and \emph{Mechanical Vibration}.
There is no general concept of \enquote{Radiation Tolerance} in Wikidata at the time this evaluation was conducted\footnote{Since Wikidata is updated regularly, the results from OE can be different when new relevant concepts are added to Wikidata.}. 
This is why we need the domain specific ontologies as well.
However, most technical entries in data sheets refer to more common concepts, e.g. mass and temperature, so they were correctly matched.

We had a domain expert adding the correct entities (group III) to the enriched ontology and disjointing the incorrect entities (group II) from the enriched ontology.
Then, in the second iteration, we increased the number of data sheets, i.e. using the data sheets Corpus B, and run the DKE and the OE again.
The result from this iteration shows a significant improvement, since the classes in the group III from the first iteration became a part of the group I 
In addition, the incorrect entities were disjointed from the classes (group II) so that the misleading concepts are excluded from these classes in the future. 

After the third iteration, using the data sheets Corpus C, most of the classes in the core ontology are enriched correctly. 
The class \emph{Mechanical Vibration} is now enriched with the entity\emph{Vibratrion} which was discovered after the second iteration. 
This shows that the ontology can be progressively improved with more data sheets and the feedback from domain experts.

\subsubsection{Star Tracker Ontology: }
%Additionally, we applied the ontology enhancement process to the star tracker ontology.
Here, the OE considered only the intrinsic classes defined within the ontology without considering imported classes from other ontologies like the core ontology.
In the star-tracker ontology, 29 classes were enriched with the entities retrieved from Wikidata.
Similar to the previous evaluation, we executed the DKE and the OE on the data sheets Corpus A, then we had a domain expert adjusting the ontology.
Afterwards, the second iteration was repeated on the data sheets Corpus B and C respectively.

%\input{tables/OntologyEnricher-result2.tex}
%\begin{figure}
%\includegraphics[width=\columnwidth]{images/OntologyEnricher-result2.png}
%\centering
%\caption{Summary of the classes in the enriched star tracker ontology. 
%OE discovered relevant entities for 19 classes.
%Thus, recall is $17/19=0.895$, and precision is $17/29=0.586$.
%F-measure is therefore 0.708.}
%\label{fig:ont-enriched-star-tracker}
%\end{figure}

Also shown in Table \ref{tab:ont-enriched}, the number of correctly enriched classes increases with respect to the number of iteration.
However, at the end of the third iteration, the classes in the group IV is still the same as the first iteration.
Compared to the core ontology, classes defined for the star tracker ontology are more specific to the domain. 
Thus, the chance to find the relevant entities is expected to be lower.
The classes that yielded no relevant entities (group IV) are: \emph{Attitude Accuracy XY and Z}, \emph{Pixel Size X and Y}, \emph{Single Star Accuracy}, \emph{Single Star Accuracy Bias}, \emph{Single Star Accuracy Noise}, and \emph{SNR}.

We notice that all classes, for which OE failed to find the relevant entities, have compound names. 
These class names should be analyzed further to find the main feature.
For example, \emph{Single Star Accuracy Noise} has \enquote{Noise} as a main feature, and \enquote{Accuracy Noise} as an extension of the main feature.

\subsection{Information Extractor} 
We expected that the enriched ontologies allow the IE to retrieve more information from the given data sheets. 
In this evaluation, we first show how an iterative OE process improves the IE. 
Then, we compare the IE using the enriched ontologies with a manual search for properties key-value on star tracker data sheets.

We used the IE to extract property-value pairs from 16 star tracker data sheets.
The property names were derived from the core and the star tracker ontologies, we then detected the property-value pairs from the data sheets.
As shown in Table \ref{tab:enhanced-ie-onto}, the enriched ontologies increase the number of discovered property-value pairs.

\begin{table}[h]\small
  \centering
  \renewcommand{\arraystretch}{1.2}
  \setlength\tabcolsep{6pt}
%\resizebox{\textwidth}{!}{%
\begin{tabular}{ c | c | c | c | c }
\hline  

\multicolumn{1}{c|}{ \begin{tabular}[c]{@{}c@{}} Ontology\end{tabular}}
& \begin{tabular}[c]{@{}c@{}} Initial\\ontology\end{tabular} 
& \begin{tabular}[c]{@{}c@{}} After \\ iteration 1\end{tabular} 
& \begin{tabular}[c]{@{}c@{}} After \\ iteration 2\end{tabular} 
& \begin{tabular}[c]{@{}c@{}} After \\ iteration 3\end{tabular} \\
\hline
	\rowcolor{gray!20} 
	Core 
	& 162  & 170 & 306 & 427\\

	Star Tracker 
	& 150 & 165 & 285 & 318\\

\hline
\end{tabular}\\
%}
\caption{Number of property-value pairs extracted from star tracker data sheets by the IE based on the initial ontologies, enriched ontologies from the iteration one, two, and three.} \label{tab:enhanced-ie-onto}
\end{table}

However, the number alone cannot assure the quality of the results.
Therefore, we had an expert manually assessed the ideal results.
We also compared our results with the baseline, which is using a text-based search over the data sheets.
A comparison in Table \ref{tab:enhanced-ie} shows that the precision of extracted values needs to be improved, while the recall is very satisfactory. 
Although the text-based search method yields better recall, but it loses its precision in exchange.
In practice, even the text-based search can cover more results, it is undesirable when one half of the return results are irrelevant. 

\begin{table}[h]\small
\renewcommand{\arraystretch}{1.3}
\setlength\tabcolsep{6pt}
\resizebox{\textwidth}{!}{%
\begin{tabular}{ c | c | c c c | c c c }
\hline
&
& \multicolumn{3}{c|}{ \begin{tabular}[c]{@{}c@{}}
	Extracted property-value pairs \end{tabular} } 
& \multicolumn{1}{c}{ }
& \multicolumn{1}{c}{ }
& \multicolumn{1}{c}{ }   \\ \cline{3-5}

\multicolumn{1}{c|}{ \multirow{-2}{*}{Ontology}} 
&\multicolumn{1}{c|}{ \multirow{-2}{*}{\begin{tabular}[c]{@{}c@{}} Extraction\\approach\end{tabular}}} 
& \multicolumn{1}{c|}{ Ideal }  
& \multicolumn{1}{c|}{ Extracted }  
& \multicolumn{1}{c|}{\begin{tabular}[c]{@{}c@{}} Correctly \\extracted\end{tabular}} 
& \multicolumn{1}{c}{ \multirow{-2}{*}{Recall}}
& \multicolumn{1}{c}{ \multirow{-2}{*}{Precision}}  
& \multicolumn{1}{c}{ \multirow{-2}{*}{F-measure}} \\
\hline
	 \cellcolor{gray!20} 
	& \cellcolor{gray!20}ConTrOn  & \cellcolor{gray!20}  
	& \cellcolor{gray!20}427 & \cellcolor{gray!20}256
	& \cellcolor{gray!20}0.94 &\cellcolor{gray!20}0.6 & \cellcolor{gray!20}0.73 \\

	\multicolumn{1}{c|}{\cellcolor{gray!20}\multirow{-2}{*}{\begin{tabular}[c]{@{}c@{}} Core\end{tabular}}}
	& \begin{tabular}[c]{@{}c@{}} Text-based\\search \end{tabular} 
	& \cellcolor{gray!20}\multirow{-2}{*}{271}
	& 542 & 263 & 0.97 & 0.49 & 0.65\\
\hline

	& \cellcolor{gray!20}ConTrOn &  
	& \cellcolor{gray!20}318 & \cellcolor{gray!20}130 
	& \cellcolor{gray!20}0.81  & \cellcolor{gray!20}0.41 & \cellcolor{gray!20}0.54\\
	
	\multicolumn{1}{c|}{ \multirow{-2}{*}{\begin{tabular}[c]{@{}c@{}} Star Tracker\end{tabular}}} 
	& \begin{tabular}[c]{@{}c@{}}  Text-based\\search \end{tabular} 
	& \multirow{-2}{*}{160}
	& 750 & 154 & 0.96 & 0.21 & 0.34\\
	
\hline
\end{tabular}
}\\

\caption{Number of property-value pairs extracted from star tracker data sheets by the IE compared between using ConTrOn and text-based search. The ontology used in this evaluation is the result from the OE after the third iteration.} \label{tab:enhanced-ie}
\end{table}

It is also worth to note that the IE performs worse with the star tracker ontology than with the core ontology. 
This is because the properties derived from the star tracker ontology are very specific and the OE could not find all information in the generic knowledge base (see Table \ref{tab:ont-enriched}).

\section{Conclusion \& Future Work}\label{sec:conclusion}
In this paper, we presented ConTrOn, a system that automatically enriches ontologies with information extracted from data sheets and semantic knowledge bases.
%We implemented three submodules: the Domain Knowledge Extractor (DKE), the Ontology Enricher (OE), and the Information Extractor (IE).
%The evaluation results show that the enriched ontology can improve the ontology-based IE process.
%Although the output from OE, which is an input of IE, is not satisfactory (F-measure of 0.654), IE yielded F-measure of 0.8. 
%This reflects that IE can even perform better if we improve the OE module.
We implemented a proof of concept that showed promising results in the evaluation demonstrating that an enriched ontology can indeed improve the information extraction.

Besides improving on precision and recall for each component, we are looking forward to the following improvements. 
First, IE currently performs the text-based search around the keywords for corresponding values.
An extraction of table structure using technique from \cite{Chao2004,Ramakrishnan2012} may capture more information from PDF files.

Second, IE is guided by class \enquote{labels}, \enquote{alternate labels}, and \enquote{common categories} from ontologies. 
We can include more class attributes such as \enquote{same as}, \enquote{defined by}, or \enquote{superclass} to retrieve more relevant results.
However, the extracted result should not be treated equally, but rather given a weight defining the level of relevancy, e.g., the term corresponding to a class attribute \enquote{same as} should be prioritized over an attribute \enquote{superclass}.
%According to the results from IE, most of the data sheets use the short form of terms, such as \emph{interface}, \emph{data format}, or \emph{output format}.
%Although it seems that these terms can be found by omitting some words, IE needs to understand if the shortened terms still retain their meaning.

Third, the OE process is semi-automatic and requires a human judgment. 
Therefore, a graphical user interface (GUI) is needed to assist users in the ontology enrichment process. 
Via this GUI, users can collaborate to improve the correctness of the information extraction. 

Furthermore, we are currently using Wikidata and WordNet as knowledge bases. 
We will include more data sources, such as DBpedia \cite{BIZER2009}, to gain more information.
Although we evaluated ConTrOn using a spacecraft development use case, its design is not domain-specific and thus we are confident that it can be applied also to other disciplines, but this is something to look into in the future.
A topic close to spacecraft components is that of electronic components and especially those that are space qualified, so this is a promising candidate to look into next.

\section*{Acknowledgement}
We wish to acknowledge the support by Philipp~Matthias~Sch{\"a}fer and Laura~Thiele from the German Space Center, DLR, Institute of Data Science as well as the advice by Philipp Martin Fischer and Andreas Gerndt from DLR Simulation and Software Technology.

\bibliographystyle{splncs04}
\bibliography{ISWC2019} 

\begin{thebibliography}{10}
\providecommand{\url}[1]{\texttt{#1}}
\providecommand{\urlprefix}{URL }
\providecommand{\doi}[1]{https://doi.org/#1}

\bibitem{Agirre2000}
Agirre, E., Ansa, O., Hovy, E.H., Mart\'{i}nez, D.: Enriching very large
  ontologies using the {WWW}. CoRR  \textbf{cs.CL/0010026} (2000),
  \url{http://arxiv.org/abs/cs.CL/0010026}

\bibitem{Asim2018}
Asim, M.N., Wasim, M., Khan, M.U.G., Mahmood, W., Abbasi, H.M.: {A survey of
  ontology learning techniques and applications}. Database  \textbf{2018}
  (October 2018). \doi{10.1093/database/bay101}

\bibitem{Baclawski2017}
Baclawski, K., Bennett, M., Berg-Cross, G., Fritzsche, D.M., Schneider, T.,
  Sharma, R., Sriram, R.D., Westerinen, A.: Ontology summit 2017 communiqu\'{e}
  - ai, learning, reasoning and ontologies. Applied Ontology  \textbf{13},
  3--18 (2017). \doi{10.3233/ao-170191}

\bibitem{Barkschat2014}
Barkschat, K.: Semantic information extraction on domain specific data sheets.
  In: ESWC (2014). \doi{10.1007/978-3-319-07443-6\_60}

\bibitem{BIZER2009}
Bizer, C., Lehmann, J., Kobilarov, G., Auer, S., Becker, C., Cyganiak, R.,
  Hellmann, S.: Dbpedia - a crystallization point for the web of data. Journal
  of Web Semantics  \textbf{7}(3),  154 -- 165 (2009).
  \doi{10.1016/j.websem.2009.07.002}, the Web of Data

\bibitem{Bozsak2002}
Bozsak, E., Ehrig, M., Handschuh, S., Hotho, A., Maedche, A., Motik, B.,
  Oberle, D., Schmitz, C., Staab, S., Stojanovic, L., Stojanovic, N., Studer,
  R., Stumme, G., Sure-Vetter, Y., Tane, J., Volz, R., Zacharias, V.: Kaon -
  towards a large scale semantic web. In: EC-Web (2002).
  \doi{10.1007/3-540-45705-4\_32}

\bibitem{Chao2004}
Chao, H., Fan, J.: Layout and content extraction for pdf documents. In:
  Document Analysis Systems. pp. 213--224. Springer Berlin Heidelberg (2004).
  \doi{10.1007/978-3-540-28640-0\_20}

\bibitem{contron2019}
ConTrOn: Contron - spacecraft parts ontology (Apr 2019).
  \doi{10.5281/zenodo.2620892}

\bibitem{Dal2012}
Dal, A., Maria, J.: Simple method for ontology automatic extraction from
  documents. International Journal of Advanced Computer Science and
  Applications  \textbf{3}(12) (2012). \doi{10.14569/ijacsa.2012.031206}

\bibitem{Fellbaum2000}
Fellbaum, C.: Wordnet : an electronic lexical database (2000).
  \doi{10.2307/417141}

\bibitem{Fischer2017}
Fischer, P.M., L\"{u}dtke, D., Lange, C., Roshani, F.C., Dannemann, F., Gerndt,
  A.: {Implementing Model-Based System Engineering for the Whole Lifecycle of a
  Spacecraft}. {CEAS} Space Journal  \textbf{9}(3),  351--365 (Jul 2017).
  \doi{10.1007/s12567-017-0166-4}

\bibitem{Maedche2000}
Maedche, A., Maedche, E., Staab, S.: The text-to-onto ontology learning
  environment. In: Software Demonstration at ICCS-2000 - Eight International
  Conference on Conceptual Structures (2000)

\bibitem{Maynard2008}
Maynard, D., Li, Y., Peters, W.: {NLP Techniques for Term Extraction and
  Ontology Population}. In: Ontology Learning and Population (2008)

\bibitem{Maynard2006}
Maynard, D., Peters, W., Li, Y.: Metrics for evaluation of ontology-based
  information extraction. In: In WWW 2006 Workshop on Evaluation of Ontologies
  for the Web (2006)

\bibitem{Moldovan2000}
Moldovan, D.I., Girju, R.: Domain-specific knowledge acquisition and
  classification using wordnet. In: Proceedings of the Thirteenth International
  Florida Artificial Intelligence Research Society Conference. pp. 224--228.
  AAAI Press (2000), \url{http://dl.acm.org/citation.cfm?id=646813.707655}

\bibitem{Murdaca2018}
Murdaca, F., Berquand, A., Kumar, K., Riccardi, A., Soares, T., Geren\'{e}, S.,
  Brauer, N.: Knowledge-based information extraction from datasheets of space
  parts. In: 8th International Systems \& Concurrent Engineering for Space
  Applications Conference (September 2018)

\bibitem{Oro2008}
Oro, E., Ruffolo, M.: {XONTO}: An ontology-based system for semantic
  information extraction from {PDF} documents. 2008 20th IEEE International
  Conference on Tools with Artificial Intelligence  \textbf{1},  118--125
  (2008). \doi{10.1109/ictai.2008.48}

\bibitem{Petasis2011}
Petasis, G., Karkaletsis, V., Paliouras, G., Krithara, A., Zavitsanos, E.:
  Ontology Population and Enrichment: State of the Art, pp. 134--166. Springer
  Berlin Heidelberg, Berlin, Heidelberg (2011).
  \doi{10.1007/978-3-642-20795-2\_6}

\bibitem{DP2019}
Peters, D., Fischer, P.M., Schäfer, P.M., Opasjumruskit, K., Gerndt, A.:
  Digital availability of supplier information for collaborative engineering of
  spacecraft. In: The International Conference on Cooperative Design,
  Visualization and Engineering (2019), \url{https://arxiv.org/abs/1905.12548}

\bibitem{Ramakrishnan2012}
Ramakrishnan, C., Patnia, A., Hovy, E.H., Burns, G.A.P.C.: Layout-aware text
  extraction from full-text pdf of scientific articles. In: Source Code for
  Biology and Medicine. vol.~7. Springer Nature (May 2012).
  \doi{10.1186/1751-0473-7-7}

\bibitem{Rizvi2018}
Rizvi, S.T.R., Mercier, D., Agne, S., Erkel, S., Dengel, A., Ahmed, S.:
  Ontology-based information extraction from technical documents. In:
  Proceedings of the 10th International Conference on Agents and Artificial
  Intelligence. {SCITEPRESS} - Science and Technology Publications (2018).
  \doi{10.5220/0006596604930500}

\bibitem{Vrandecic2014}
Vrande\v{c}i\'{c}, D., Kr\"{o}tzsch, M.: Wikidata: A free collaborative
  knowledgebase. Commun. ACM  \textbf{57}(10),  78--85 (Sep 2014).
  \doi{10.1145/2629489}

\bibitem{Wimalasuriya2010}
Wimalasuriya, D.C., Dou, D.: Ontology-based information extraction: An
  introduction and a survey of current approaches. Journal of Information
  Science  \textbf{36},  306--323 (2010). \doi{10.1177/0165551509360123}

\bibitem{Wu1994}
Wu, Z., Palmer, M.: Verbs semantics and lexical selection. In: Proceedings of
  the 32Nd Annual Meeting on Association for Computational Linguistics. pp.
  133--138. ACL '94, Association for Computational Linguistics, Stroudsburg,
  PA, USA (1994). \doi{10.3115/981732.981751}

\end{thebibliography}
\end{document}